\documentclass{article}
\usepackage{amsfonts}
\usepackage{amsmath}

\newtheorem{theorem}{Theorem}

\newtheorem{proposition}[theorem]{Proposition}
\newtheorem{remark}[theorem]{Remark}

\input{tcilatex}
\begin{document}

\title{On a new unified geometric description of gravity and electromagnetism%
}
\author{Nicoleta VOICU \\
"Transilvania" University, Brasov, Romania}
\date{}
\maketitle

\begin{abstract}
In a previous paper, we have introduced a new unified description of the
main equations of the gravitational and of the electromagnetic field, in
terms of tidal tensors and connections on the tangent bundle $TM$ of the
space-time manifold. In the present work, we relate these equations to
variational procedures on the tangent bundle. The Ricci scalar of the
proposed connection is dynamically equivalent to the usual Einstein-Maxwell
Lagrangian. Also, in order to be able to perform these variational
procedures, we find an appropriate completion of the metric tensor (from the
base manifold) up to a metric structure on $TM.$
\end{abstract}



\section{Introduction}

The main request of a unified, geometric description of gravity and
electromagnetism is to find some geometric structures such that, in the
Einstein field equations, the electromagnetic stress-energy tensor is
enclosed in the left hand side, i.e., in the Einstein tensor. In the period
between the two world wars, remarkable theories were built by: Einstein
(teleparallelism, metric-affine theories), Eddington, Eisenhart, Schr\H{o}%
dinger (affine geometry with torsion), Weyl (gauge theory), Kaluza and Klein
(fifth space-time dimension). Afterwards, the interest for the classical
approaches somehow waned and the focus shifted rather to quantum theories, 
\cite{Goenner}.

More recently, interest for classical unified theories has grown again.
Thus, Ferraris and Kijowski, \cite{Ferraris}, Chrusciel, \cite{Chrusciel},
Poplawski, \cite{Poplawski1}, developed the purely affine approach, in which
the torsion of the affine connection (on the space-time manifold) accounts
for electromagnetism.

A second path -- which we follow here -- uses the geometry of the tangent
bundle $TM$ of the space-time manifold. Here, we should first mention the
description due to R. Miron and collaborators, \cite{Lagrange}, \cite%
{Miron-Buchner}, \cite{Ingarden}, \cite{Miron-Balkan}, in which the metric
tensor characterizes gravity, while electromagnetism is encoded in
connections on the tangent bundle\textit{.} In the cited papers, they obtain
a geometrization of the Lorentz equations of motion of charged particles and
of Maxwell equations -- but they do not solve the problem of enclosing the
Maxwell stress-energy tensor into the Einstein tensor.

Other theories on $TM$ try to include information regarding electromagnetism
in Finsler-type metrics (Randers, Beil or Weyl metrics, \cite{Beil}, \cite%
{Beil2}). Also, recently, Wanas, Youssef and Sid-Ahmed produced another
description, \cite{Wanas}, based on teleparallelism on $TM$. A version using
complex Lagrange geometry is proposed by G. Munteanu, \cite{Munteanu}.

In \cite{Voicu-tidal},\ we proposed a new approach -- based, on one side, on
the notion of geodesic deviation (and subsequently, of tidal tensor, \cite%
{Costa}) and on the other side, on Miron's idea of encoding the information
regarding gravity in the metric tensor on the base manifold and
electromagnetism, in a 1-parameter family of connections $(\overset{\alpha }{%
N},\overset{\alpha }{D})$ (where $\overset{\alpha }{N}$ are Ehresmann
connections and $\overset{\alpha }{D},$ affine connections) on $TM.$ We
chose $(\overset{\alpha }{N},\overset{\alpha }{D})$ such that:

1) worldlines of charged particles define autoparallel curves for both $%
\overset{\alpha }{N},$ $\overset{\alpha }{D}$;

2)\ worldline deviation equations are as simple as possible (their right
hand side does not depend on the derivatives of the deviation vector field);

3) the Ricci tensor\ of a connection $\overset{\alpha }{D}$ can be obtained
just by differentiating the trace of the tidal tensor with respect to the
fiber coordinates on $TM$.

In \cite{Voicu-tidal}, we wrote Maxwell equations directly in terms of tidal
tensors attached to $\overset{\alpha }{N}$.

In the present paper, we build an analogue of the classical Hilbert action
based on Ricci scalars of $\overset{\alpha }{D}$.

But, in order to perform variational procedures on $TM$, we also need a
volume element on $TM.$ With this aim, we propose a completion of the
Lorentzian metric $g_{ij}$ up to a metric on the total space $TM,$ with two
properties: a) there exists, for each $x\in M,$ a canonical domain of
integration $\tilde{\Delta}\subset T_{x}M$ with respect to the fiber
coordinates such that, for functions $f=f(x)$ defined on $M$, the integral
of $f$ on a domain $\Delta \subset M$ and the integral of $f$ on $\Delta
\times \tilde{\Delta}$ coincide; b) the divergence of the horizontal lift to 
$TM$ of a vector field on $M$ coincides with the divergence of the original
vector field.

This construction refines the one in \cite{Finsler-electro}. With this, we
get one more property of the connections $\overset{\alpha }{D}:$

4) for a conveniently chosen $\alpha ,$ the Ricci scalar of $\overset{\alpha 
}{D}$ is dynamically equivalent to the usual Einstein-Maxwell Lagrangian on $%
M$. Einstein field equations (with the electromagnetic stress-energy tensor
included in the Einstein tensor) and stress-energy conservation are then
obtained in the usual way.

Property 4)\ is similar to the one in Kaluza-Klein theory, but it does not
require additional space-time dimensions; moreover, our method has the
advantage of providing geometrizations of the Lorentz equations of motion
and of worldline deviation equations.

\section{Preliminaries}

\subsection{Basic equations}

Consider a 4-dimensional, $\mathcal{C}^{\infty }$ Lorentzian manifold $(M,g)$%
, with local coordinates $(x^{i})_{i=\overline{0,3}},$ regarded as
space-time manifold and $\nabla ,$ its Levi-Civita connection, with
coefficients $\gamma _{~jk}^{i}$ and curvature tensor $r_{j~kl}^{~i};$ we
denote by $\partial _{i}$ the elements of the natural basis for the module
of vector fields on $M$.

In general relativity, the metric components $g_{ij}$ describe the
gravitational field. The electromagnetic field is described by the potential
1-form:%
\begin{equation}
A=A_{i}(x)dx^{i}  \label{def_A}
\end{equation}%
and by the electromagnetic field tensor (Faraday 2-form) $F=dA,$ or,
locally, 
\begin{equation}
F=\dfrac{1}{2}F_{ij}dx^{i}\wedge dx^{j},~\ \ \ F_{ij}=\nabla
_{i}A_{j}-\nabla _{j}A_{i}.  \label{F_local}
\end{equation}%
From the definition of $F$, it follows the identity: $dF=0,$ which is
equivalent to homogeneous Maxwell equations: $\nabla _{\partial
_{i}}F_{jk}+\nabla _{\partial _{k}}F_{ij}+\nabla _{\partial _{j}}F_{ki}=0.$

\bigskip

The other basic equations of the two physical fields are obtained by
variational methods. The total action attached to these, together with a
system of particles with masses $m_{a},$ coordinates $x_{a}^{i}$ and
electric charges $q_{a}$ is, \cite{Landau}:

\begin{eqnarray}
S &=&-\underset{S_{m}}{\underbrace{\sum m_{a}c\int ds}}-\underset{S_{i}}{%
\underbrace{\sum \dfrac{q_{a}}{c}\int A_{k}(x)dx^{k}}}-
\label{general_Lagrangian} \\
&&-\underset{S_{f}}{\underbrace{\dfrac{1}{16\pi c}\int F_{ij}F^{ij}\sqrt{-g}%
d^{4}x}}-\underset{S_{g}}{\underbrace{\dfrac{c^{3}}{16\pi k}\int r\sqrt{-g}%
d^{4}x}};  \notag
\end{eqnarray}%
where the sums are taken over the particles in the system, $g=\det (g_{ij}),$
$d^{4}x=dx^{0}\wedge dx^{1}\wedge dx^{2}\wedge dx^{3},$ $r$ is the Ricci
scalar of $g$ and $c,k$ are constants (the light speed in vacuum and the
gravitational constant). The volume integrals are taken over a large enough
compact domain $\Delta \subset M$\footnote{%
Traditionally, one integrates over a finite amount of time and over the
whole spatial manifold, under the assumption that at infinity, the fields
vanish. Loosely speaking, we can integrate over a large enough compact
region of $M.$}. The first term $S_{p}$ characterizes free particles, the
third one $S_{f}$ characterizes the electromagnetic field and the second one 
$S_{i}$, the interaction between the field and the particles. The fourth
integral $S_{g}$ is the Hilbert action for $g_{ij}$.

The line integrals $S_{m}$ and $S_{i}$ can be transformed into volume
integrals, by means of the Dirac delta function (involving the variables $%
x^{1},x^{2},x^{3}$):%
\begin{eqnarray}
S_{m} &=&-\int \mathcal{L}_{m}\sqrt{-g}d^{4}x,~\ \ \mathcal{L}_{m}=\underset{%
a}{\sum }\dfrac{||dx||}{dx_{a}^{0}}\dfrac{m_{a}c}{\sqrt{-g}}\delta
^{3}(x-x_{a}),  \label{S_m} \\
S_{i} &=&-\dfrac{1}{c^{2}}\int A_{i}J^{i}\sqrt{-g}d^{4}x,~~J^{i}=\underset{a}%
{\sum }\dfrac{dx_{a}^{i}}{dx_{a}^{0}}\dfrac{q_{a}c}{\sqrt{-g}}\delta
^{3}(x-x_{a}).  \label{S_i}
\end{eqnarray}

Here, $\mathcal{L}_{m}$ is a scalar, while $J^{i}$ are components of the 
\textit{4-current vector field, }\cite{Landau}. Thus, the total action $S$
becomes:%
\begin{equation}
S=-\int (\mathcal{L}_{m}+\dfrac{1}{c^{2}}A_{i}J^{i}+\dfrac{1}{16\pi c}%
F_{ij}F^{ij}+\dfrac{c^{3}}{16\pi k}r)\sqrt{-g}d^{4}x.  \label{S_simplified}
\end{equation}

\bigskip

\textbf{I. }Variation of the 4-potential $A$ in $S$\ yields the \textit{%
inhomogeneous Maxwell equations}: 
\begin{equation}
\nabla _{\partial _{j}}F^{ij}=-\dfrac{4\pi }{c}J^{i}.  \label{inhom_max}
\end{equation}

\textbf{II. }Variation of $S$ with respect to the metric components $g^{ij}$%
\ leads, \cite{Landau}, \cite{Bertschinger}, to the \textit{Einstein field
equations:}%
\begin{equation}
G_{ij}=\dfrac{8\pi k}{c^{4}}T_{ij};  \label{E-Max_classical}
\end{equation}%
here, the Einstein tensor%
\begin{equation*}
G_{ij}=r_{ij}-\dfrac{1}{2}rg_{ij}
\end{equation*}%
is obtained from: $\delta S_{g}=-\dfrac{c^{3}}{16\pi k}\int G_{ij}\delta
g^{ij}\sqrt{-g}d^{4}x=\dfrac{c^{3}}{16\pi k}\int G^{ij}\delta g_{ij}\sqrt{-g}%
d^{4}x$.

In the right hand side of (\ref{E-Max_classical}), the stress-energy tensor $%
T_{ij}$ is written as a sum: 
\begin{equation*}
T_{ij}=~\overset{~~f}{T}_{ij}+\overset{m}{T}_{ij},
\end{equation*}

where:

- $\overset{~~f}{T}_{ij}$is the stress-energy tensor of the electromagnetic
field%
\begin{equation*}
\overset{~~f}{T}_{ij}=\dfrac{1}{4\pi }(-F_{il}F_{j}^{~~l}+\dfrac{1}{4}%
g_{ij}F^{lm}F_{lm}),
\end{equation*}%
obtained from: $\ \ \delta S_{f}=\dfrac{1}{2c}\int \overset{~~f}{T}%
_{ij}\delta g^{ij}\sqrt{-g}d^{4}x=-\dfrac{1}{2c}\int \overset{~~f}{T}\overset%
{}{^{ij}}\delta g_{ij}\sqrt{-g}d^{4}x$;

- $\overset{m}{T}_{ij}$ is the stress-energy tensor of matter:

$\delta S_{m}=\dfrac{1}{2c}\int \overset{m}{T}_{ij}\delta g^{ij}\sqrt{-g}%
d^{4}x=-\dfrac{1}{2c}\int \overset{m}{T}\overset{}{^{ij}}\delta g_{ij}\sqrt{%
-g}d^{4}x$.

\bigskip

\textbf{III. }In the case of a single particle (with mass $m$ and charge $q$%
), variation of $S$ with respect to its trajectory, i.e., the variation of:%
\begin{equation}
S_{m}+S_{i}:=mc\int ds+\dfrac{q}{c}\int A_{k}(x)dx^{k}
\label{particle_action}
\end{equation}%
with respect to $x^{i}=x^{i}(s)$ (where $s$ is the arc length), leads to the 
\textit{Lorentz equations of motion of charged particles:}%
\begin{equation}
\dfrac{\nabla \dot{x}^{i}}{ds}=\dfrac{q}{mc^{2}}F_{~j}^{i}\dot{x}^{j}.
\label{Lorentz_eq_classical}
\end{equation}

\subsection{Conservation laws}

Since action $S$ is a scalar, it is invariant to diffeomorphisms, \cite%
{Bertschinger}. This invariance leads to the well-known energy-momentum
conservation law.

Consider diffeomorphisms with pushforward $\tilde{x}^{i}=x^{i}+\varepsilon
\xi ^{i}(x)$ on $M$ (where $\xi ^{i}$ are components of a vector field and $%
\varepsilon >0$). Then, the variations of the field variables are given by
their Lie derivatives:%
\begin{equation}
\delta A_{i}=\mathcal{L}_{\xi }A_{i}=\xi ^{k}\nabla _{\partial
_{k}}A_{i}+A_{k}\nabla _{\partial _{i}}\xi ^{k},~\ \delta g_{ij}=\mathcal{L}%
_{\xi }g_{ij}=\nabla _{\partial _{j}}\xi _{i}+\nabla _{\partial _{i}}\xi
_{j}.  \label{Lie_derivs}
\end{equation}

The variation of the total action is $\delta S=\delta _{A}S+\delta _{g}S;$
in detail: 
\begin{equation*}
\delta S=\int \left\{ -\dfrac{1}{4\pi c}(\nabla _{\partial _{j}}F^{ij}+%
\dfrac{4\pi }{c}J^{i})\delta A_{i}+\dfrac{c^{3}}{16\pi k}(G^{ij}-\dfrac{8\pi
k}{c^{4}}T^{ij})\delta g_{ij}\right\} \sqrt{-g}d^{4}x.
\end{equation*}

For solutions $F$ of the inhomogeneous Maxwell equations (\ref{inhom_max}),
it remains%
\begin{equation*}
\delta S=\dfrac{c^{3}}{16\pi k}\int (G^{ij}-\dfrac{8\pi k}{c^{4}}%
T^{ij})\delta g_{ij}\sqrt{-g}d^{4}x
\end{equation*}%
Substituting $\delta g_{ij}$ from (\ref{Lie_derivs}) and integrating by
parts, we get:%
\begin{equation*}
\delta S=\dfrac{c^{3}}{8\pi k}\int {\Large \{}\nabla _{\partial
_{j}}[(G^{ij}-\dfrac{8\pi k}{c^{4}}T^{ij})\xi _{i}]-\xi _{i}\nabla
_{\partial _{j}}(G^{ij}-\dfrac{8\pi k}{c^{4}}T^{ij}){\Large \}}\sqrt{-g}%
d^{4}x.
\end{equation*}

The first term becomes, by Stokes' theorem, a boundary one, hence it will
not contribute to the integral; we get:%
\begin{equation*}
\delta S=-\dfrac{c^{3}}{8\pi k}\int \xi _{i}\nabla _{\partial _{j}}(G^{ij}-%
\dfrac{8\pi k}{c^{4}}T^{ij}){\Large \}}\sqrt{-g}d^{4}x.
\end{equation*}%
Since the variations $\xi _{i}$ are independent, we are led to:%
\begin{equation*}
\nabla _{\partial _{j}}(G^{ij}-\dfrac{8\pi k}{c^{4}}T^{ij})=0.
\end{equation*}

Contracted Bianchi identities tell us that $\nabla _{\partial _{j}}G^{ij}=0$%
. We thus get the \textit{energy-momentum conservation law}%
\begin{equation}
\nabla _{\partial _{j}}T^{ij}=0.  \label{e_m_conservation}
\end{equation}%
In more detail, this is \cite{Landau}: $\nabla _{\partial _{j}}\overset{~f}{T%
}\overset{}{^{ij}}=-\dfrac{1}{c}F_{~j}^{i}J^{j}=-\nabla _{\partial _{j}}%
\overset{~m}{T}\overset{}{^{ij}}$).

\section{Geometric structures on $TM$}

\subsection{Ehresmann connections}

Consider now the tangent bundle $(TM,\pi ,M),$ with local coordinates $%
(x\circ \pi ,y)=:(x^{i},y^{i})_{i=\overline{0,3}};$ we denote by%
\begin{equation}
l=\dfrac{y}{\left\Vert y\right\Vert },~~\left\Vert y\right\Vert =\sqrt{%
g_{ij}y^{i}y^{j}},  \label{supporting element}
\end{equation}%
the \textit{normalized supporting element} on $TM$, \cite{Shen}, and by $%
_{,i}$ and $\cdot _{i},$ partial differentiation with respect to $x^{i}$ and 
$y^{i}$ respectively. An Ehresmann connection $N$ on $TM,$ \cite{Lagrange}, 
\cite{Shen}, gives rise to the adapted basis 
\begin{equation}
(\delta _{i}=\dfrac{\partial }{\partial x^{i}}-N_{~i}^{j}(x,y)\dfrac{%
\partial }{\partial y^{j}},~~\ \ \dot{\delta}_{i}=\dfrac{\partial }{\partial
y^{i}}),  \label{general_adapted_basis}
\end{equation}%
and to its dual $(dx^{i},\delta y^{i}=dy^{i}+N_{~j}^{i}dx^{j}).$

Consider the following 1-parameter family of \textit{Randers-type
Lagrangians,} \cite{Shen}, \cite{Ingarden} depending on a real parameter $%
\alpha :$%
\begin{equation}
\overset{\alpha }{L}~=\sqrt{g_{ij}(x)\dot{x}^{i}\dot{x}^{j}}+\alpha A_{i}(x)%
\dot{x}^{i}.  \label{randers}
\end{equation}

The action $\int \overset{\alpha }{L}dt$ attached to $\overset{\alpha }{L}$
is formally similar to the action (\ref{particle_action}); though using the
same notations as in the previous section, for the moment, we will not
attribute any physical significance to $A$ or $\alpha $. Taking $%
t=const\cdot s$ as a parameter, extremal curves $x=x(t)$ are given by: 
\begin{equation}
\dfrac{dy^{i}}{dt}+\gamma _{~jk}^{i}y^{j}y^{k}-\alpha \left\Vert
y\right\Vert F_{~j}^{i}y^{j}=0,~\ y^{i}=\dot{x}^{i},  \label{Lorentz_eq}
\end{equation}%
where%
\begin{equation}
F_{~j}^{i}:=g^{ih}(\nabla _{\partial _{h}}A_{j}-\nabla _{\partial
_{j}}A_{h}),~\ \ \ \left\Vert y\right\Vert =\sqrt{g_{ij}y^{i}y^{j}}.
\label{F_ij}
\end{equation}%
We obtain a 1-parameter family of sprays, \cite{Anto}, \cite{Lagrange}, $G=%
\overset{\alpha }{G}$ on $TM:$%
\begin{equation}
2\overset{\alpha }{G}\overset{}{^{i}}(x,y)=\gamma _{~jk}^{i}y^{j}y^{k}+2%
\overset{\alpha }{B}\overset{}{^{i}},  \label{spray_randers}
\end{equation}%
with%
\begin{equation}
2\overset{\alpha }{B}\overset{}{^{i}}=-\alpha \left\Vert y\right\Vert
F_{~j}^{i}y^{j}=:-\alpha \left\Vert y\right\Vert F^{i};  \label{definition_B}
\end{equation}%
the corresponding spray connections, \cite{Anto}, $N=\overset{\alpha }{N}$
have the coefficients:%
\begin{equation}
\overset{\alpha }{G}\overset{}{_{~j}^{i}}:=\overset{\alpha }{G}\overset{}{%
_{~\cdot j}^{i}}=\gamma _{~jk}^{i}y^{k}+B_{~j}^{i}.
\label{spray_conn_coeffs}
\end{equation}

If there is no risk of confusion upon $\alpha $, we will denote simply $%
G^{i},B^{i},\delta _{i},G_{~j}^{i},B_{~j}^{i}...$ instead of $\overset{%
\alpha }{G}\overset{}{^{i}},$ $\overset{\alpha }{B}\overset{}{^{i}},$ $%
\overset{\alpha }{\delta }_{i}\overset{\alpha }{G}\overset{}{_{~j}^{i}},$ $%
\overset{\alpha }{B}\overset{}{_{~j}^{i}}$ etc.

Extremal curves of the action $\int \overset{\alpha }{L}dt$ \ are thus
autoparallel curves (\textit{geodesics)} for $N=\overset{\alpha }{N}:$%
\begin{equation*}
\dfrac{\delta y^{i}}{dt}=0,~~y^{i}=\dot{x}^{i},~\ i=\overline{0,3}
\end{equation*}%
and geodesic deviations are given, \cite{Voicu-tidal}, by:%
\begin{equation}
\dfrac{\delta ^{2}w^{i}}{dt^{2}}=E_{~j}^{i}w^{j},~\ \
~~~~~E_{~j}^{i}=R_{~jk}^{i}y^{k},\   \label{deviations_N_tidal}
\end{equation}%
where $R_{~jk}^{i}=\delta _{k}N_{~j}^{i}-\delta _{j}N_{~k}^{i}$ are the
local coefficients of the curvature of $N.$

We will call the quantity%
\begin{equation}
E~\mathbb{=~}E_{~j}^{i}\delta _{i}\otimes dx^{j},  \label{tidal_tensor}
\end{equation}%
the \textit{tidal tensor\footnote{%
The tidal tensor is tightly related to the J\textit{acobi endomorphism }$%
\Phi $\textit{\ }in \cite{BCD}. }} associated to $N.$

The functions $B^{i}$ in (\ref{definition_B}) are the components of a
horizontal vector field $B=B^{i}\delta _{i}$ on $TM.$ Their derivatives with
respect to the fiber coordinates are:%
\begin{equation}
B_{~j}^{i}=B_{~\cdot j}^{i}=-\dfrac{\alpha }{2}(F^{i}l_{j}+\left\Vert
y\right\Vert F_{~j}^{i}),~\ B_{~jk}^{i}:=B_{~\cdot jk}^{i}=-\dfrac{\alpha }{2%
}(l_{\cdot jk}F^{i}+l_{j}F_{~k}^{i}+l_{k}F_{~j}^{i}).  \label{derivs_B}
\end{equation}

Conversely, from the homogeneity of degree 2 of $B$ in the fiber
coordinates, it follows: $B_{~j}^{i}y^{j}=2B^{i},~\ \
B_{~jk}^{i}y^{k}=B_{~j}^{i}~~$etc.

\subsection{Affine connections on $TM$}

Consider%
\begin{equation}
G_{~jk}^{i}:=G_{~\cdot jk}^{i}=\gamma _{~jk}^{i}+B_{~jk}^{i};
\label{Berwald_conn}
\end{equation}%
and the affine connections $D=\overset{\alpha }{D}$ on $TM$ which act on the 
$\overset{\alpha }{N}$-adapted basis vectors as: 
\begin{equation}
D_{\delta _{k}}\delta _{j}=G_{jk}^{i}\delta _{i},~\ \ D_{\delta _{k}}\dot{%
\delta}_{j}=G_{jk}^{i}\dot{\delta}_{i},~\ \ \ D_{\dot{\delta}_{k}}\delta
_{j}=D_{\dot{\delta}_{k}}\dot{\delta}_{j}=0.  \label{d-connection}
\end{equation}

Connections $\overset{\alpha }{D},$ $\alpha \in \mathbb{R},$ preserve by
parallelism the distributions generated by $\overset{\alpha }{N}$ (hence,
they are \textit{distinguished linear connections, }\cite{Lagrange}, on $TM$%
), i.e., for any two vector fields $X,Y$ on $TM,$ we have: $%
D_{X}(hY)=hD_{X}Y,$ $D_{X}(vY)=vD_{X}Y.$ They are, generally, non-metrical.

\bigskip

$D=\overset{\alpha }{D}$ has generally nonvanishing torsion, given by:%
\begin{equation}
\mathbb{T}=R_{~jk}^{i}\dot{\delta}_{i}\otimes dx^{j}\otimes dx^{k}
\label{torsion_D}
\end{equation}%
and the curvature of $D$ is:\textit{\ }%
\begin{eqnarray}
\mathbb{R} &=&R_{j~kl}^{~i}\delta _{i}\otimes dx^{j}\otimes dx^{k}\otimes
dx^{l}+~R_{j~kl}^{~i}\dot{\delta}_{i}\otimes \delta y^{j}\otimes
dx^{k}\otimes dx^{l}+  \notag \\
&&+B_{j~kl}^{~i}\delta _{i}\otimes dx^{j}\otimes dx^{k}\otimes \delta y^{l},
\label{curvature_D}
\end{eqnarray}%
where $B_{j~kl}^{~i}=B_{\cdot jkl}^{i}$ and $R_{j~kl}^{~i}$ are obtained in
terms of the tidal tensor as: 
\begin{equation}
R_{j~kl}^{~i}=\dfrac{1}{2}(E_{~k}^{i})_{\cdot jl}~.
\label{curvature_components}
\end{equation}%
In particular, the Ricci tensor of $\overset{\alpha }{D}$ is obtained from
the trace $E_{~i}^{i}$: 
\begin{equation}
R_{jl}=-\dfrac{1}{2}(E_{~i}^{i})_{\cdot jl}=R_{j~li}^{~i}.
\label{Ricci tensor}
\end{equation}

Conversely, the tidal tensor $E$ can be written\footnote{%
Expression (\ref{relation_E_R}) points out an almost complete similarity
between the tidal tensor and the notion of flag curvature in Finsler
geometry. The difference consists in the metric tensor used in raising and
lowering indices, which is here $g_{ij}$ (not the Finslerian one
corresponding to $\overset{\alpha }{L}$)\ and which leads to somehow
different properties.} in terms of $\mathbb{R}$ as:%
\begin{equation}
E_{~k}^{i}=R_{j~kl}^{~i}y^{j}y^{l},~\ E_{~i}^{i}=-R_{jl}y^{j}y^{l}.
\label{relation_E_R}
\end{equation}

\bigskip

\textbf{Particular case: }For $\alpha =0,$ we get: $2\overset{0}{G}\overset{}%
{^{i}}=\gamma _{~jk}^{i}y^{j}y^{k},$ i.e.,%
\begin{equation*}
\overset{0}{G}\overset{}{_{~jk}^{i}}=\gamma _{~jk}^{i};
\end{equation*}%
for vector fields $X,Y\ $on $M,$ we have $l_{h}(\nabla _{X}Y)=\overset{0}{D}%
_{l_{h}(X)}l_{h}(Y)$ (where $l_{h}$ denotes the horizontal lift to $TM$);
thus, $\overset{0}{D}$ can be considered as the $TM$-equivalent of the
Levi-Civita connection $\nabla $ and each $\overset{\alpha }{D},$ as a
"perturbation\footnote{%
Each of these perturbations gives rise to a notion of product for vector
fields on $TM;$ thus, the module of vector fields on $TM$ becomes an algebra
-- the so-called \textit{deformation algebra, } \cite{Nicolescu}.}" of $%
\overset{0}{D},$ with contortion tensor $B.$ We obviously have: $\overset{0}{%
E}\overset{}{_{~j}^{i}}=r_{j~kl}^{~i}y^{j}y^{l},~$ $\overset{0}{R}\overset{}{%
_{j~kl}^{~i}}=r_{j~kl}^{~i}$\ and the Ricci tensor of $\overset{0}{D}$ is $%
\overset{0}{R}\overset{}{_{jk}}=r_{jk}.$

In \cite{Voicu-tidal}, we have proved that the Euler-Lagrange equations for $%
\overset{\alpha }{L}$ are equivalent to: 
\begin{equation}
D_{V}V=0.
\end{equation}%
where $V$ is the complete lift of the velocity vector field $\dot{x}%
^{i}\partial _{i}:$%
\begin{equation}
V:=y^{i}\delta _{i}+\dfrac{\delta y^{i}}{dt}\dot{\delta}_{i},~\ y^{i}=\dfrac{%
dx^{i}}{dt}  \label{V}
\end{equation}%
and geodesic deviations can be also written as%
\begin{equation}
\dfrac{D^{2}w^{i}}{dt^{2}}=E_{~k}^{i}w^{k};  \label{geodesic_deviation_tidal}
\end{equation}%
here, all covariant derivatives are considered "with reference vector $y$", 
\cite{Shen}, i.e., in their local expressions, $G_{~j}^{i}=G_{~j}^{i}(x,y),$ 
$G_{~jk}^{i}=G_{~jk}^{i}(x,y).$

\subsection{Metric structure on $TM$}

Fix a connection $\overset{\alpha }{N}.$ The Lorentzian metric $g=(g_{ij})$
on $M$ can be lifted into a metric%
\begin{equation*}
g=g_{ij}(x)dx^{i}\otimes dx^{j}
\end{equation*}%
on the horizontal subbundle of $TM,$ which we will extend up to a metric on
the whole $TM:$ 
\begin{equation*}
\mathcal{G}:=g_{ij}(x)dx^{i}\otimes dx^{j}+v_{ij}(x)\delta y^{i}\otimes
\delta y^{j}.
\end{equation*}%
\bigskip

Let us consider $v_{ij}$ with the following properties: 1) $v_{ij}$ -
positive definite (Riemannian) and 2) the determinants of $(g_{ij})$ and $%
(v_{ij})$ have equal absolute values.

\bigskip

Such a choice is always possible. For instance, in Riemann normal
coordinates $(x^{i^{\prime }})$ for $g$ at some $x_{0}\in M$ (i.e., $%
g_{i^{\prime }j^{\prime }}=\eta _{i^{\prime }j^{\prime }}=diag(-1,1,1,1)$),
we can set: $v_{i^{\prime }j^{\prime }}(x_{0})=\delta _{i^{\prime }j^{\prime
}},$ i.e., in the adapted basis, 
\begin{equation*}
\mathcal{G}(x_{0}^{\prime }):=diag(-1,1,1....,1)
\end{equation*}%
(in another coordinate system $(x^{k})$, we will have $v_{kl}=\dfrac{%
\partial x^{i^{\prime }}}{\partial x^{k}}\dfrac{\partial x^{j^{\prime }}}{%
\partial x^{l}}v_{i^{\prime }j^{\prime }}$). With this choice, the
dependence $x\mapsto \mathcal{G}(x)$ (accordingly, $x\mapsto v_{ij}(x)$) is
a smooth one and 
\begin{equation*}
v:=\det (v_{ij})=\det (\dfrac{\partial x^{i^{\prime }}}{\partial x^{k}}%
)^{2}=-g.
\end{equation*}

\bigskip

As a consequence, the volume element on $TM$ is:%
\begin{equation}
d\Omega =\sqrt{-gv}d^{4}x\wedge \delta ^{4}y,  \label{volume_el_rough}
\end{equation}%
where $\delta ^{4}y=\delta y^{0}\wedge \delta y^{1}\wedge \delta y^{2}\wedge
\delta y^{3}.$ Moreover, $d^{4}x\wedge \delta y^{i}=d^{4}x\wedge
(dy^{i}+N_{~j}^{i}dx^{j})=d^{4}x\wedge dy^{i},$ i.e., we can actually write:%
\begin{equation}
d\Omega =\sqrt{-gv}d^{4}x\wedge d^{4}y.  \label{volume_el_refined}
\end{equation}

Since $v$ is positive definite, the set: 
\begin{equation}
\tilde{\Delta}=\{y\in T_{x}M~|~v_{ij}y^{i}y^{j}\leq \sqrt[4]{2/\pi ^{2}}%
\},~\ \ r>0,~x\in M  \label{delta_tilda}
\end{equation}%
is a compact subset of $T_{x}M;$ in normal coordinates for $v_{ij}$, the
domain $\tilde{\Delta}=\tilde{\Delta}(x)$ actually becomes a 3-sphere of
volume equal to 1 in 4-dimensional Euclidean space.

For any $x\in M,$ we will set $\tilde{\Delta}$ as a canonical integration
domain with respect to $y\in T_{x}M;$ thus, for any compact domain $\Delta
\subset M$ and for any function $f:\Delta \rightarrow \mathbb{R}$, the
integral of $f$ over\footnote{%
For the sake of simplicity, we denote by the same letter $f$ the composition 
$f\circ \pi .$} $\Delta \times \tilde{\Delta}\subset TM$ coincides with its
integral over $\Delta :$ 
\begin{equation}
\underset{\Delta \times \tilde{\Delta}}{\int }f(x)d\Omega =\underset{\Delta }%
{\int }f(x)\cdot vol(\tilde{\Delta})\sqrt{-g}d^{4}x=~\underset{\Delta }{\int 
}f(x)\sqrt{-g}d^{4}x.  \label{integral_f(x)}
\end{equation}

\bigskip

The divergence of a horizontal vector field $X=X^{i}(x,y)\delta _{i}$ on $TM$
is, \cite{Finsler-electro}:%
\begin{eqnarray*}
&&div(X)=\dfrac{1}{\sqrt{-gv}}\delta _{i}(X^{i}\sqrt{-gv})-N_{~i\cdot
j}^{j}X^{i}= \\
&=&\delta _{i}X^{i}+X^{i}\delta _{i}(\ln \sqrt{-g})+X^{i}\delta _{i}(\ln 
\sqrt{v})-N_{~i\cdot j}^{j}X^{i}.
\end{eqnarray*}

From (\ref{derivs_B}), it follows: $B_{~i\cdot j}^{j}=0,$ that is,$%
~N_{~i\cdot j}^{j}=\gamma _{~ij}^{j}.$ Taking into account that $\delta
_{i}(\ln \sqrt{-g})=\delta _{i}(\ln \sqrt{v})=\gamma _{~ij}^{j},$ we get: 
\begin{equation}
div(X)=\dfrac{1}{\sqrt{-g}}\delta _{i}(X^{i}\sqrt{-g})=~\overset{0}{D}%
_{\delta _{i}}X^{i}.  \label{divergence_X}
\end{equation}

In particular, for a vector field $Y=Y^{i}(x)\partial _{i}$ on the base
manifold, we have: $\ div(Y)=~\overset{0}{D}_{\delta _{i}}Y^{i}=div(l_{h}Y)$.

\section{Einstein field equations}

\subsection{In vacuum}

Consider $\alpha $ as arbitrary and fixed. As analogue of the classical
Hilbert action (this time, involving both $g_{ij}$ and $F_{ij}$), we propose:%
\begin{equation*}
S_{fg}=-\dfrac{c^{3}}{16\pi k}\underset{\Delta \times \tilde{\Delta}}{\int }%
Rd\Omega ,
\end{equation*}%
where $R$ is the Ricci scalar of $\overset{\alpha }{D}$ with $\tilde{\Delta}$
as in (\ref{delta_tilda}) and $\Delta $ as in Section 2.

From (\ref{Ricci tensor}), we get, by direct computation: 
\begin{equation}
R=r+~\overset{0}{D}_{\delta _{i}}(B_{~~j}^{ij})-\dfrac{1}{2}%
g^{jk}(B_{~h}^{i}B_{~i}^{h})_{\cdot jk};
\end{equation}%
the term $\overset{0}{D}_{\delta _{i}}(B_{~~j}^{ij})\ $is a divergence,
i.e., it will only produce a boundary term, which finally vanishes. A brief
calculation leads to $-\dfrac{1}{2}g^{jk}(B_{~h}^{i}B_{~i}^{h})_{\cdot jk}=%
\dfrac{3\alpha ^{2}}{2}F_{ij}F^{ij};$ thus, the two remaining terms in the
integral $S_{fg}$ do not depend on $y$ any longer, i.e., $S_{fg}$ can
finally be written as an integral on the base manifold. We thus get:

\begin{theorem}
The Ricci scalar $R$ of $D=\overset{\alpha }{D}$ it is dynamically
equivalent to the following Lagrangian on $M$: 
\begin{equation}
\tilde{R}=r+\dfrac{3\alpha ^{2}}{2}F_{ij}F^{ij}.  \label{R_tilda}
\end{equation}
\end{theorem}

In particular, for $\alpha =\alpha ^{\ast }$ given by: 
\begin{equation}
\dfrac{3(\alpha ^{\ast })^{2}}{2}=\dfrac{k}{c^{4}},  \label{special_alpha}
\end{equation}%
we get the usual Einstein-Maxwell action: 
\begin{equation*}
S_{fg}(\alpha ^{\ast })=-\dfrac{c^{3}}{16\pi k}\int \tilde{R}\sqrt{-g}%
d^{4}x=-\dfrac{c^{3}}{16\pi k}\int (r+\dfrac{k}{c^{4}}F_{ij}F^{ij})\sqrt{-g}%
d^{4}x.
\end{equation*}

In terms of $\overset{\mathbf{\alpha }}{D}$, we get, by a similar procedure
to the one in \cite{Landau}:

\begin{equation*}
\delta _{g}S_{fg}(\alpha ^{\ast })=-\dfrac{c^{3}}{16\pi k}\int (\tilde{R}%
_{ij}-\dfrac{1}{2}\tilde{R}g_{ij}+\mathcal{B}_{\cdot ij})\delta g^{ij}\sqrt{%
-g}d^{4}x,
\end{equation*}%
where: $\mathcal{B}:=\dfrac{3}{2}\dfrac{B^{l}B_{l}}{\left\Vert y\right\Vert
^{2}}+\dfrac{1}{2}B_{~h}^{i}B_{~i}^{h}.$ Thus:

\begin{proposition}
\label{EFE_vacuum}Einstein-Maxwell equations in vacuum are expressed in
terms of $D=D(\alpha ^{\ast })$ as:%
\begin{equation}
\mathcal{G}_{ij}=0,  \label{E-M_vacuum}
\end{equation}%
where 
\begin{equation}
\mathcal{G}_{ij}=\tilde{R}_{ij}-\dfrac{1}{2}\tilde{R}g_{ij}+\mathcal{B}%
_{\cdot ij}.  \label{E_tensor_vacuum}
\end{equation}%
are components of a symmetric horizontal tensor field $\mathcal{G}=\mathcal{G%
}_{ij}dx^{i}\otimes dx^{j}$ on $TM.$
\end{proposition}

\subsection{In the presence of matter}

Assuming that we also have some particles of masses $m_{a}$ and electric
charges $q_{a}$, the total action is: 
\begin{equation}
S=S_{fg}+S_{m}+S_{i}.  \label{total_action_TM}
\end{equation}

\begin{remark}
The sum $S_{mi}:=S_{m}+S_{i}$ is written in terms of the functions $\overset{%
\alpha }{L}$ as:%
\begin{equation}
S_{mi}=-\int \underset{a}{\sum }\mu c\overset{\alpha _{a}}{L}\sqrt{-g}%
d^{4}x,~\ \ \ \ \ \ \mu :=\dfrac{m_{a}\delta ^{3}(x-x_{a})}{\sqrt{-g}},
\end{equation}%
where, for each particle, we have a different value of $\alpha :$%
\begin{equation*}
\alpha _{a}=\dfrac{q_{a}}{mc^{2}}.
\end{equation*}
\end{remark}

The total action is, then:%
\begin{equation}
S=-\int (\dfrac{c^{3}}{16\pi k}\tilde{R}(\alpha ^{\ast })+\underset{a}{\sum }%
\mu cL_{(\alpha _{a})})\sqrt{-g}d^{4}x.  \label{total_action_M}
\end{equation}

The action $S$ in (\ref{total_action_M})\ is nothing but the usual total
action in Section 2.1, but here, the term $S_{f}$ corresponding to the
electromagnetic field is contained in the Ricci scalar $R$ (equivalently, in 
$\tilde{R}$).

\bigskip

By varying the expression (\ref{total_action_TM}) of $S$ with respect to $%
g^{ij}$, and using the fact that actually, $S_{i}$ does not depend on $%
g^{ij},$ we get, similarly to Proposition \ref{EFE_vacuum}:

\begin{proposition}
Einstein field equations are written as:%
\begin{equation}
\mathcal{G}_{ij}=\dfrac{8\pi k}{c^{4}}\overset{m}{T}_{ij},
\label{einstein-maxwell}
\end{equation}%
where $\mathcal{G}_{jk}$ is the generalized Einstein tensor (\ref%
{E_tensor_vacuum}) (including the electromagnetic part of the stress-energy
tensor) and $\overset{m}{T}_{ij}$ is the stress-energy tensor of matter.
\end{proposition}

\section{Invariance to diffeomorphisms and conservation laws}

Since we have proven that the total action $S=S_{fg}+S_{m}+S_{i}$ is
equivalent to an action on the base manifold, it is enough to consider
diffeomorphisms of $M,$ as in Section 2.2. By a similar reasoning, we get
that, as long as inhomogeneous Maxwell equations are satisfied by $A$, there
hold the equalities:%
\begin{equation*}
\nabla _{\partial _{j}}(\mathcal{G}^{ij}-\dfrac{8\pi k}{c^{4}}\overset{m}{T}%
\overset{}{^{ij}})=0,
\end{equation*}%
which is read on $TM$ as:%
\begin{equation}
div(\mathcal{G}-\dfrac{8\pi k}{c^{4}}\overset{m}{T})=0.
\label{conservation_law}
\end{equation}

The above is just the usual energy-momentum conservation law, expressed in
terms of the generalized Einstein tensor\footnote{%
Here, the terms $div(\mathcal{G})$ and $div(\overset{m}{T})$ are generally,
not separately conserved.} $\mathcal{G}_{ij}.$

Relation (\ref{conservation_law}) is a consequence of the Bianchi identity
for the horizontal component of the curvature and of Maxwell equations. The
rest of the Bianchi identities for $\overset{\alpha }{D}$ do not yield any
new information (the "perturbation" terms appearing from $F$ cancel one
another).

\section{Equations of motion of charged particles}

For a single charged particle, subject to the gravitational and the
electromagnetic fields, we have:%
\begin{equation*}
S_{mi}=-mc\int \overset{\alpha }{L}(x,\dot{x})dt,
\end{equation*}%
with $\alpha =\dfrac{q}{mc^{2}}.$ The equations of motion are:%
\begin{equation}
\dfrac{\overset{\alpha }{D}y^{i}}{dt}=0,~\ y^{i}=\dot{x}^{i}~\ (\alpha =%
\dfrac{q}{mc^{2}});  \label{Lorentz_eqns}
\end{equation}%
(where $t=const\cdot s$); for particles having the same ratio $\dfrac{q}{m},$
worldline deviation equations are given by 
\begin{equation}
\dfrac{\overset{\alpha }{D}\overset{}{^{2}}w^{i}}{dt^{2}}=\overset{\alpha }{E%
}\overset{}{_{~j}^{i}}w^{j},~~\alpha =\dfrac{q}{mc^{2}}.
\label{worldline_deviation}
\end{equation}%
with $E$ as in (\ref{tidal_tensor}).

\textbf{Acknowledgment. }The work was supported by the Sectorial Operational
Program Human Resources Development (SOP HRD), financed from the European
Social Fund and by the Romanian Government under the Project number
POSDRU/89/1.5/S/59323.

\end{document}